\documentclass[prc,showpacs,preprintnumbers,
               amsmath,amssymb,twocolumn,
               superscriptaddress]{revtex4}               
\usepackage{graphicx} 
\usepackage{dcolumn}  
\usepackage{bm}       

\usepackage{epsfig}
\begin{document}

\title{Density matrix renormalization group approach to two-fluid open many-fermion systems}

\author{J. Rotureau}
\affiliation{Department of Physics and Astronomy, University of Tennessee,
Knoxville, Tennessee 37996}
\affiliation{Physics Division, Oak Ridge National Laboratory, Oak Ridge, Tennessee 37831}
\affiliation{Joint Institute for Heavy Ion Research,
Oak Ridge National Laboratory, P.O. Box 2008, Oak Ridge, Tennessee 37831}
\author{N. Michel}
\affiliation{Department of Physics, Graduate School of Science,
Kyoto University, Kitashirakawa, Kyoto, 606-8502, Japan}
\affiliation{CEA, Centre de Saclay, IRFU/Service de Physique 
Nucl\'eaire, F-91191 Gif-sur-Yvette, France}
\author{W. Nazarewicz}
\affiliation{Department of Physics and Astronomy, University of Tennessee,
Knoxville, Tennessee 37996}
\affiliation{Physics Division, Oak Ridge National Laboratory, Oak Ridge, Tennessee 37831}
\affiliation{Institute of Theoretical Physics, University of Warsaw, ul. Ho\.za 69, 00-681 Warsaw, Poland}
\author{M. P{\l}oszajczak}
\affiliation{Grand Acc\'el\'erateur National d'Ions Lourds (GANIL), CEA/DSM - CNRS/IN2P3, BP 55027,
F-14076 Caen Cedex, France}
\author{J. Dukelsky}
\affiliation{Instituto de Estructura de la Materia, CSIC, Serrano 123, 28006 Madrid, Spain}

\date{\today}

\begin{abstract}
We have extended  the density matrix renormalization group (DMRG)
approach to two-fluid open many-fermion systems governed by 
complex-symmetric Hamiltonians. The applications are carried out for
three- and four-nucleon (proton-neutron) systems within   the Gamow
Shell Model (GSM) in the complex-energy plane. We study necessary and
sufficient conditions for the GSM+DMRG method to yield the correct ground
state eigenvalue and discuss different truncation schemes within DMRG.
The proposed approach will enable  configuration interaction  studies of
weakly-bound and unbound strongly interacting complex systems which,
because of a prohibitively large size of Fock space,  cannot be treated
by means of the direct  diagonalization.

\end{abstract}


\maketitle

\section{Introduction}

The theoretical description of weakly bound and unbound states in
atomic nuclei requires a rigorous treatment
of   many-body correlations in the presence of  scattering continuum
and decay channels \cite{Jac_d,Jac_o,ppnp}. Many-body states located close to the particle 
emission threshold display unusual properties
such as, e.g., halo and Borromean structures, clusterization phenomena,  and 
cusps in various observables resulting from the strong 
coupling to the continuum space. These peculiar
features cannot be described in the standard shell model (SM) in which the
single-particle (s.p.) basis is usually derived from an infinite well such as the harmonic
oscillator potential. Indeed, resonance and scattering states, which play a
decisive role in the structure of weakly bound/unbound states, are not properly treated
within the standard SM formalism.

The solution of the configuration-interaction problem in the presence of continuum
states has been recently
advanced in the real-energy continuum SM \cite{Kar,Rot,Volya}  and
in the complex-energy SM, the so-called Gamow Shell Model (GSM)
\cite{Mic02a,Bet02,Bet04,Mic06,Hag06}. In the GSM, the s.p. basis is given
by the Berggren ensemble containing Gamow states and the non-resonant
continuum of scattering states. The scattering states are distributed along a
contour defined in the complex $k$-plane and,
together  with the Gamow states, form a complete set \cite{Berggren}.
In practice, the contour is
discretized  and the many-body  basis is
spanned by the Berggren ensemble.
As in standard SM applications, the dimension of the many-body valence
space increases dramatically with the number of valence nucleons and the size
of the s.p. basis. Moreover, the use of the Berggren ensemble implies
complex-symmetric matrices for the representation of the Hermitian
Hamilton operator. Consequently, efficient numerical methods are
needed to solve the many-body Schr\"{o}dinger equation of GSM. The
DMRG approach is ideally suited to
optimize the size of the scattering space in the GSM problem as the properties
of the non-resonant shells vary smoothly along the scattering contour.

The DMRG method was first introduced to overcome the limitations of the
Wilson-type renormalization group to describe strongly correlated 1D
lattice  systems with short-range interactions \cite{dmrg1} (see recent
reviews \cite{Rev1,Rev2,Rev3}). More recently, by reformulating the DMRG
in a s.p. basis, several applications  to finite Fermi systems like
molecules \cite{pap2,chan}, superconducting grains \cite{pap3,delft}, quantum dots
\cite{pap4,weiss},  atomic nuclei \cite{pap1}, and fractional quantum Hall systems \cite{Fei} 
 have been reported. While
most of the DMRG studies were focused on equilibrium properties in
strongly correlated closed quantum systems characterized by  Hermitian
density matrices,  non-equilibrium systems involving non-Hermitian and
non-symmetric density matrices can also be treated \cite{pap6}. Nuclear
applications of  DMRG  in the context of the standard SM, both in the
$M$-scheme and in the angular-momentum conserving $J$-scheme, have also been
reported \cite{pap5,pitsan,pitsan1} with mixed success.
In  the previous study \cite{Rot2}, we reported the first
application of the DMRG method in
the context of the GSM and showed that in this case the  method
provides a highly accurate description of broad resonances
 in the neutron-rich nuclei with few valence particles. %

The present study is an extension of  the previous work \cite{Rot2} to
the case where both protons and neutrons are included in the valence
space. Several significant improvements over Ref.~ \cite{Rot2} have been
made concerning the DMRG algorithm and numerical implementation.
Our work is organized as follows. In
Sec.~\ref{sec_GSM}, we briefly recapitulate the GSM formalism and the generalized variational
principle behind it. Section~\ref{sec_dm} describes the DMRG method in
the $J$-scheme  as applied to the open-system formalism of the GSM. Practical applications of the GSM+DMRG method
are presented  in Sec.~\ref{Applications}. Illustrative calculations are carried out for $^7$Li (three-nucleon
systems) and  $^8$Li (four-nucleon systems). We study the impact of different starting conditions on the DMRG
result and introduce  two different truncation schemes in the DMRG procedure and discuss their virtues.  The
truncation schemes are compared with the GSM benchmark diagonalization results. The resulting efficient
calculation  scheme  opens a window for extending calculations to systems  beyond the current limits of direct
diagonalization. The conclusions of our work are contained in Sec.~\ref{sec_sum}.
\section{The Gamow Shell Model} \label{sec_GSM}
\begin{figure}[hbt]
\begin{center}
\includegraphics[width=0.4\textwidth]{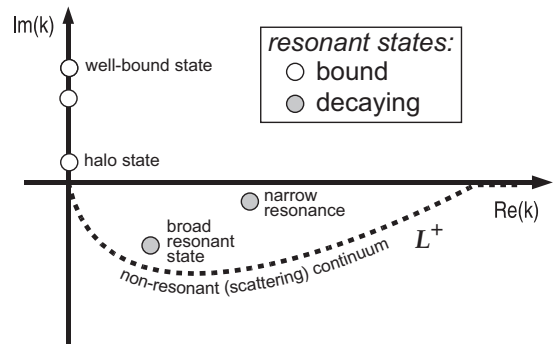}
\caption{Illustration of the Berggren completeness relation
(\protect\ref{Ber_com})
in the complex $k$-plane. The bound states are located on the positive imaginary
axis. The weakly bound halo states lie close to the origin. The positive-energy resonant
states are located in the fourth quadrant. Those with a small imaginary part
can be interpreted as resonances.
The complex-$k$ shells
on the contour $L^{+}$ represent the  non-resonant scattering continuum.}
\label{k_plane}
\end{center}
\end{figure}

The s.p. basis used in the GSM formalism is generated by a finite depth potential. It
forms a complete set of states
in the sense of the Berggren completeness relation \cite{Berggren}:
\begin{eqnarray}
\sum_{n=b,d}|\tilde{u}_n\rangle \langle u_n|+\int_{L^{+}}|\tilde{u}_{k}\rangle \langle u_{k}|=1,
\label{Ber_com}
\end{eqnarray}
where the discrete sum includes s.p. bound (negative-energy)
resonant states ($b$) and positive-energy decaying resonant states ($d$).
These states are the poles of the corresponding one-body
scattering matrix.
The integration in Eq.~(\ref{Ber_com}) is performed along a contour
$L^{+}$ defined in the complex $k$-plane that is  located {\it below}
the resonant states included in the basis (see Fig. \ref{k_plane}). In
general, different
contours can be used for each
$\ell,j$ partial wave.
By discretizing continuum states on $L^{+}$ a finite s.p. basis is obtained.
The many-body basis is obtained in the usual way by constructing
product states (Slater determinants)
from this discrete s.p. set. We assume in the following
that the nucleus can be described as a
system of $n_{\pi}$ protons and $n_{\nu}$ neutrons evolving around a
closed core. Within this picture, the GSM Hamiltonian $\hat{H}$ reads:
\begin{eqnarray}
\hat{H} &=&\sum_{i=1}^{n_{\pi}+n_{\nu}}
\left[ \frac{ {\bf{{p_{i}}^2}}}{2\mu}+ U_{i}\right]+\sum_{j>i=1}^{n_{\pi}+n_{\nu}}V_{ij},
\label{hamiltonian}
\end{eqnarray}
where ${\bf{{p_{i}^2}}}/{2\mu}$ is the s.p. kinetic energy operator,
$\mu$ is the reduced mass of the nucleon+core system,
$U_{i}$ is the finite-depth, one-body
potential, and $V_{ij}$ is the two-body residual interaction.

The GSM Hamiltonian  (\ref{hamiltonian}) is complex symmetric.  According to the
complex variational principle \cite{moisey1,moisey}, a complex analog to the usual variational principle
for self-conjugated Hamiltonians, the Rayleigh quotient
\begin{eqnarray}
E[\Phi]=\frac{\langle \Phi^{*}|\hat{H}|\Phi\rangle}{\langle \Phi^{*}| \Phi \rangle}
\end{eqnarray}
 is stationary around any eigenstate $|\Phi_0\rangle$ of $\hat{H}$:
\begin{equation}\label{eigenv}
 \hat{H}|\Phi_0\rangle=E[\Phi_{0}]|\Phi_0\rangle.
\end{equation} 
 That is, at
$|\Phi\rangle$=$|\Phi_0\rangle$
 the variation of the functional $E[\Phi]$ is zero:
\begin{equation}\label{genvar}
\delta_\Phi{E[\Phi]}_{\Phi=\Phi_0}=0.
\end{equation}
It should be noted \cite{moisey1,moisey} that the
 complex variational principle is  a stationary principle rather than an upper or
lower bound for either the real or imaginary part of the complex eigenvalue.
However, it can be very useful when applied to the squared modulus of the complex eigenvalue \cite{moisey2}.
Indeed, 
\begin{equation}\label{genvar1}
\delta_\Phi{|E|^2}=\delta_\Phi{(E^*E)}=E^*\delta_\Phi{E}+E\delta_\Phi{E^*}=0 
\end{equation}
at $|\Phi\rangle$=$|\Phi_0\rangle$ because of analyticity of $E[\Phi]$.

\section{The Density Matrix Renormalization Group method for the Gamow Shell
Model} \label {sec_dm} Let us consider a nucleus with $n_{\pi}$ active protons and $n_{\nu}$ active neutrons and
let us denote by $|J^{\pi}\rangle$ the eigenstate of $\hat{H}$  having angular momentum $J$ and parity $\pi$. As
$|J^{\pi}\rangle$ is the many-body pole of  the scattering matrix of $\hat{H}$, the contribution from  scattering
shells on $L^{+}$ to the many-body wave function is usually smaller than the contribution from the resonant
orbits. Based on this observation, the following separation is usually performed \cite{Rot2}: the many-body states
constructed from the s.p. poles form  a subspace $A$ 
(the so-called `reference subspace'), and the remaining states
containing contributions from
 non-resonant  shells form a complement subspace
$B$. As we shall discuss  later in Sections \ref{gs7li} and
\ref{gs8li}, this intuitive  definition of the reference subspace may be insufficient for
describing certain classes of eigenstates.
In such cases, states in $A$ have to be constructed both from the s.p. poles
{\it and} selected scattering shells.

At the first stage of the GSM+DMRG method, called 'the warm-up
phase', the scattering shells  are gradually added to the reference subspace
to create the subspace $B$. This process is described in the next section.
\subsection {Warm-up phase of GSM+DMRG}
One begins by constructing all states $|k\rangle_A$  forming
 the reference
subspace $A$. The set of those states shall be denoted as $\{ k_A\}$.
The many-body configurations  in $A$
can be classified in different families  $\{n;j_A^\pi \}$
according to their number of
nucleons $n$, total angular momentum $j_A$, and parity $\pi$. In the following, we shall
omit the parity label in the
notation of a given family. States with a number of protons (neutrons) larger than
$n_{\pi}$ ($n_{\nu}$) are not
considered since they do not contribute to the many-body states in the composition
of subspaces $A$ and $B$.

All possible matrix elements of suboperators of the two-body Hamiltonian  (\ref{hamiltonian}) acting in $A$,
expressed in the second quantization form, are calculated and stored:
\begin{eqnarray}\label{suboper}
 \{O\}& =& \{a^{\dagger },
(a^{\dagger }\, \widetilde{a})^{K}, (a^{\dagger }a^{\dagger
})^{K}, ((a^{\dagger }a^{\dagger })^{K} \widetilde{a})^{L}, \nonumber \\
&& (a^{\dagger }a^{\dagger })^{K}
(\widetilde{a}\widetilde{a})^{K}\},
\end{eqnarray}
with $a^{\dagger }$ and $\widetilde{a}$  being the nucleon creation and annihilation operators in resonant shells. The GSM
Hamiltonian is then diagonalized in the reference space to provide the  zeroth-order  approximation
$|\Psi_J\rangle^{(0)}$ to  $|J^{\pi}\rangle$. This vector, called `reference state',  plays an important role in
the GSM+DMRG truncation algorithm.

In the next step,  the subspace of the
first scattering shell $(lj)_1$  belonging to the discretized contour $L^{+}$
is added.
Within this shell, one constructs all possible many-body states
$\{(lj)^{n_B}_1\}$, denoted as
$|i\rangle_B$,
grouped in  {\bf {$\{n_B;j_B\}$}} families.
Matrix elements of suboperators (\ref{suboper})
acting on  $|i\rangle_B$ are computed.
By coupling states in $A$ with the states $|i_B\rangle$, one constructs the set
$\{ k_A \otimes  i_B \}^{J}$ of states having fixed $J^{\pi}$. This ensemble serves as a basis
in which the GSM Hamiltonian is
diagonalized. Of course at this stage, the resulting
wave function is a rather poor approximation of $|J^{\pi}\rangle$
as only
one scattering shell has been
included. The target state $|\Psi_J\rangle$ is  selected among the eigenstates
of $\hat{H}$ as the one
 having the largest overlap with the reference vector $|\Psi_J\rangle^{(0)}$.
Based on  the expansion
\begin{equation}
|\Psi_J\rangle = \sum_{k_A,i_B} c^{k_A(j_A)}_{i_B(j_B)} \{ |k_A(j_A)\rangle \otimes  |i_B(j_B)\rangle \}^J,
\end{equation}
by summing over the  reference subspace $A$ for a {\em fixed} value
of $j_B$,
one defines the reduced density matrix \cite{Mcdmrg}:
\begin{eqnarray}\label{rdm}
\rho ^{B}_{i_Bi'_B}(j_B) \equiv \sum_{k_A}c^{k_A(j_A)}_{i_B(j_B)} c^{k_A(j_A)}_{i'_B(j_B)}.
\end{eqnarray}
By construction, the density matrix $\rho ^{B}$ is block-diagonal in both $j_B$.
In the warm-up
phase, the reference
subspace becomes the `medium' for the `system' part in the $B$ subspace.

Truncation in the system sector
is dictated by  the density matrix. In standard
DMRG applications for Hermitian problems where
the eigenvalues of $\rho$ are real non-negative, only
the eigenvectors corresponding to the  largest eigenvalues
are kept during the DMRG process.
Within the metric defining the Berggren ensemble, the GSM
density matrix is complex-symmetric
and its eigenvalues  are,
in general, complex. In the straightforward generalization
of the DMRG algorithm to the complex-symmetric case \cite{Rot2},
one retains  at most ${N}_{\rm opt}^{(0)}$
eigenstates of $\hat\rho^B$,
\begin{equation}\label{eigenrho}
\hat{\rho}^B (j_B)|\alpha\rangle_{B} = w_\alpha |\alpha\rangle_{B},
\end{equation}
having the largest nonzero values of  $|w_\alpha|$.
Due to the normalization of $|\Psi_J\rangle$,
the sum of all (complex) eigenvalues of $\hat\rho_B$ is equal to 1:
\begin{eqnarray}\label{feq}
Tr(\hat{\rho}^B)\equiv\sum_{\alpha}w_\alpha=1,
\end{eqnarray}
i.e., the imaginary part of the trace vanishes exactly.

Expressing the eigenstates $|\alpha\rangle_{B}$ in terms of the
vectors $|i\rangle_B$ in $B$:
\begin{equation}
|\alpha\rangle_{B}=\sum_{i} d^{\alpha}_{i}|i\rangle_B,
\end{equation}
all matrix elements of the suboperators in these optimized states,
\begin{eqnarray}
_{B}\langle \alpha|O|\beta\rangle_{B}
=\sum_{i,i'}d^{\alpha}_{i}d^{\beta}_{i'}~ _{B}\langle i|O|i'\rangle_{B},  \label{renorm}
\end{eqnarray}
are recalculated and stored.

\begin{figure}[hbt]
\begin{center}
\includegraphics[width=0.3\textwidth]{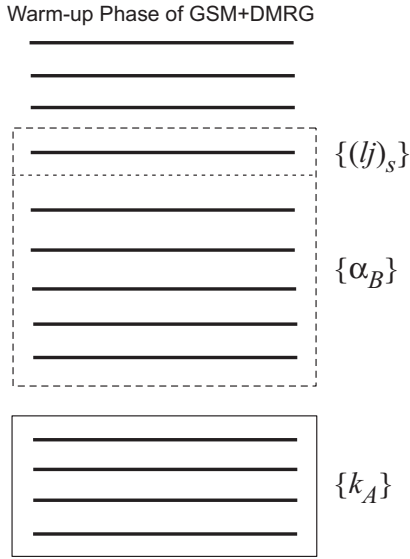}
\caption{Schematic illustration of the GSM+DMRG
procedure  during the $s^{th}$ step of the warm-up phase.
States $\{k_{A} \}$ from $A$, previously optimized states $\alpha_{B} $,
and states $\{(lj)_s \}$ constructed by occupying  the $s^{th}$ shell
with $n$ particles
are coupled to generate the
new
set of states $\{k_A \otimes i_B\}^J=
\{k_{A} \otimes  \{\alpha_B \otimes (lj)_s^n \} \}^J $.}
\label{warmup2}
\end{center}
\end{figure}
The warm-up procedure continues by adding to the system part the configurations containing
particles in the second  scattering shell $(lj)_2$.
As in the first step, one
constructs all many-body states
 $\{(lj)^n_2\}$
within this new shell and calculates corresponding
matrix elements of suboperators (\ref{suboper}).
The new vectors  $|i\rangle_B$ in the system sector are then obtained
by coupling the states
$|\alpha\rangle_{B}$ calculated in the first step with the vectors
 $|\{(lj)^n_2\}\rangle$.

Following the same prescription as before, one constructs the set
$\{k_A \otimes i_B \}^{J}$ of states coupled to $J^{\pi}$ in which the Hamiltonian
is diagonalized. As
in the previous step, the new target state for the calculation of the
reduced density matrix $\rho_B$ is defined as the one with maximum
overlap with the reference state. Again, at most ${N}_{\rm opt}^{(0)}$
eigenvectors  of $\rho_B$ are retained and all matrix elements of
suboperators for these optimized states are recalculated. This
procedure, illustrated schematically in  Fig. \ref{warmup2}, continues 
until the last shell in $B$ is reached, providing a first guess for the
wave function of the system  in the whole ensemble of shells.
At this point, all s.p. states have been considered, and all suboperators of the Hamiltonian
$\hat{H}$ acting on states
saved after truncation in $B$ have been computed and stored. The warm-up phase ends  and the
so-called sweeping phase begins.

\subsection{Sweeping phase of GSM+DMRG}

Starting from the last scattering shell $(lj)_{last}$, the procedure
continues in the reverse direction (the `sweep-down' phase) using the
previously stored information. At this stage, the meaning of the medium
and system parts changes as compared to the warm-up phase.

In the sweeping phase, the states
$|k_A\rangle$ of the reference subspace $A$ and the states $|{i_{prev}\rangle}$
generated in the warm-up phase form
the medium. The corresponding basis is:
\begin{equation}\label{newbasis}
|\Phi_{J_{part}}(k,i_{prev})\rangle =  \{|k_{A}\rangle \otimes  |i_{prev}\rangle\}^{J_{part}}.
\end{equation}
The system part is generated by adding the scattering
shells one at a time.

The sweep-down process  begins  by
constructing all possible states $|i\rangle$ from the shell
$(lj)_{last}$ and calculating the corresponding suboperators of
 $\hat{H}$. A new basis coupled to $J^{\pi}$ is then formed by
coupling states $|\Phi_{J_{part}}\rangle$
 with $|i(j_B)\rangle$.
\begin{figure}[hbt]
\begin{center}
\includegraphics[width=0.45\textwidth]{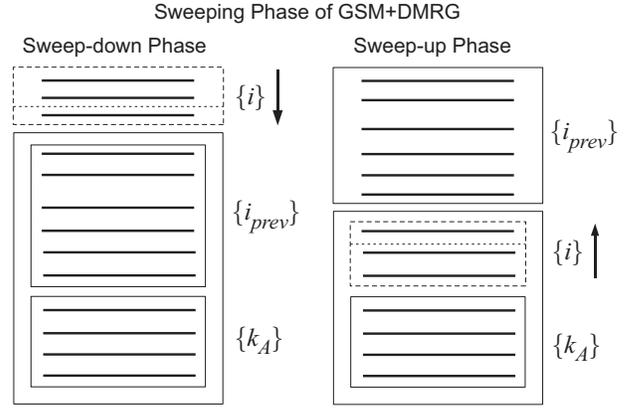}
\caption{Schematic illustration of the coupling between different
configurations  during
the sweeping phase of GSM+DMRG.
Left: sweep-down phase.
Here, configurations
$\{k_{A}\}$ are coupled with previously generated configurations
$\{i_{prev}\}$ and with $\{i\}$ states.
Right: sweep-up phase. Here, configurations $\{k_A\}$ are coupled with
$\{i\}$ and with configurations  $\{i_{prev}\}$ in $B$
generated in a previous sweep-down phase.}
\label{fig_sweep}
\end{center}
\end{figure}
The representation of  $\hat{H}$ in this basis is constructed using the
Wigner-Eckart theorem
by coupling suboperators acting in $A$,  $\{i_{prev}\}$,
and $\{i \}$ (the set of states $|i\rangle$). As before, the target state
\begin{eqnarray}
|\Psi_J\rangle=\sum_{k,i_{prev},i} c_{k,i_{prev},i}^{J_{part},j_B}\{ |\Phi_{J_{part}}(k,i_{prev})\rangle \otimes  |i (j_B)\rangle\}^{J}
\end{eqnarray}
is identified by picking up the eigenstate of $\hat{H}$ having the largest overlap with the reference state
$|\Psi_J\rangle^{(0)}$. 
The density matrix is then constructed
\begin{eqnarray}
\rho ^{B}_{ii';j_B}=\sum _{k,i_{prev},J_{part}}c^{(J_{part},j_B)}_{k,i_{prev},i}c^{(J_{part},j_B)}_{k,i_{prev},i'}
\end{eqnarray}
and diagonalized for each value of $j_{B}$. At this point, the truncation can be done
in two different ways.
In the first truncation method (i)
 at most $N_{\rm opt}$ eigenvectors of the density matrix with the largest
nonzero moduli of eigenvalues  are kept. This is precisely the truncation technique that has been employed in the
warm-up phase.  The actual number of states retained may vary since one considers only eigenvectors with nonzero
eigenvalues. The second method (ii), based on the identity (\ref{feq}), is a generalization of the dynamical block
selection approach \cite{Leg03}. Here we focus on controlling the numerical error by selecting in each step of the
procedure $\rho$, $N_{\rho}$ vectors with the largest moduli of
 the eigenvalues  so that the
 condition
\begin{eqnarray}
\left| 1-Re\left(\sum_{\alpha=1}^{N_{\rho}}w_\alpha\right)\right|< \epsilon
\label{criterion_den}
\end{eqnarray}
is satisfied. The quantity $\epsilon$ in (\ref{criterion_den}) can be viewed as
 the truncation error of the reduced density matrix.
It is worth noting that while the trace of the reduced density matrix is strictly equal to one (\ref{feq}), this
is no longer the case for the restricted sum of eigenvalues in  (\ref{criterion_den}). In particular,
 the real part of the reduced trace  may be greater than one and  the imaginary part may be nonzero.
For that reason, in Eq.~(\ref{criterion_den}) one considers  the real part of the partial trace.
 The smaller $\epsilon$,  the larger number $N_{\rho}$ of eigenvectors must be kept.
In particular, for  $\epsilon$=0,  all eigenvectors with non-zero eigenvalues are retained.
One should emphasize that $N_{\rho}$ may change from one step to another.
Section~\ref{Applications} discusses the convergence of the GSM+DMRG procedure with
respect to $N_{\rm opt}$ and $\epsilon$.

The matrix elements (\ref{renorm}) in eigenvectors
$|\alpha\rangle_{B}$ saved after the truncation
are recalculated and stored. The procedure continues by
adding the next shells one by one  until  the first scattering shell is reached.
At each step during the sweep-down
phase, all suboperators of $\hat{H}$ are stored.
The sweep-down phase of GSM+DMRG is schematically illustrated in the left portion of
Fig.~\ref{fig_sweep}.

At this point, the procedure is reversed, and a sweep in the upward direction (the `sweep-up'  phase) begins. Using
the information previously stored, a first shell is added, then a second one, etc. (see Fig. \ref{fig_sweep},
right panel). The medium now consists of states in the reference subspace $A$ and states  $\{i_{prev}\}$ in $B$
that were generated during  a previous sweep-down phase. The sweeping sequences continue until convergence for
target eigenvalue is achieved.
\section {Applications of the GSM+DMRG method} \label{Applications}

This work describes  the first GSM+DMRG treatment of  open-shell proton-neutron
nuclei. As illustrative examples, we take
$^{7}\rm{Li}$ and $^{8}\rm{Li}$ described  schematically as interacting
nucleons outside the closed core of
$^4$He. The neutron one-body potential in
Eq. (\ref{hamiltonian}) is a Woods-Saxon (WS) potential with radius $R_{0}=2$~fm
and diffuseness $d=0.65$~fm. The spin-orbit strength $V_{\rm so}=7.5$~MeV and
the depth of the central potential $V_0$=47~MeV are fixed to reproduce the
experimental energies and widths of  $3/2_1^-$
and $1/2_1^-$ resonances in $^5$He. For the protons,
the same WS average potential is supplemented by the Coulomb potential
generated by a uniformly charged sphere of radius $R_0$ and charge $Q$=+2e.

The two-body interaction in (\ref{hamiltonian}) is represented by a finite-range
Surface Gaussian Interaction (SGI) \cite {Mic04}:
\begin{eqnarray}
V_{i,j}^{J,T}& =& V_{0}(J,T)\exp \left[- \left(\frac{  {\bf{r_1}}-{\bf{r_2}} } {\mu} \right)^{2}  \right] \nonumber \\
 & \times & \delta(|{\bf{r_1}}|+|{\bf{r_2}}|-2R_0).
\end{eqnarray}
The strengths  $V_0(J,T)$ are the same as in Ref.~\cite{Mic04}.
The $T$=0 couplings  depend linearly  on
the number of valence neutrons $N_v$:
\begin{eqnarray}
V_{0}^{1,0}=\alpha_{10}-\beta_{10}(N_v-1),\nonumber \\ V_{0}^{3,0}=\alpha_{30}-\beta_{30}(N_v-1), \label{param_V}
\end{eqnarray}
where $\alpha_{10}=-600~{\rm MeV}\cdot{\rm fm}^{3}$, $\beta_{10}=-50~{\rm MeV}\cdot{\rm fm}^{3}$,
$\alpha_{30}=-625~{\rm MeV}\cdot{\rm fm}^{3}$, and $\beta_{30}=-100~{\rm MeV}\cdot{\rm fm}^{3}$.

The above  set of parameters has not been optimized to reproduce the actual structure of 
$^{7}$Li and $^{8}$Li. Our choice of interaction
is motivated by the fact that the main purpose of this study
is to test the DMRG procedure for proton-neutron systems for which the exact GSM
diagonalization is still possible. In this context, ``our" $^{7}$Li and $^{8}$Li systems
should be viewed as three- and four-nucleon cases, respectively.

Following the method described in Ref. \cite{Mic04}, s.p. bases for
protons and neutrons are generated by their respective spherical  Hartree-Fock (HF)
potentials corresponding to the GSM
Hamiltonian (\ref{hamiltonian}).  Neutron and proton
valence spaces include the $0p_{3/2}$  HF poles
as well as the scattering  shells $\{p_{3/2}\}_{\rm c}$ in the complex $k$-plane,
and the $\{p_{1/2}\}_{\rm c}$, $s_{1/2}$, and $d_{5/2}$ real-energy continua.

The $L_{\nu;p_{3/2}}^{+}$ ($L_{\pi;p_{3/2}}^{+}$) contour, along which
the scattering $\nu p_{3/2}$ ($\pi p_{3/2}$) shells are distributed,
is defined by a triangle with vertices at: $(Re(k), 
Im(k))= (0.0,0.0), (0.17,-0.1),(2,0.0)$,  and a segment along the
$Re(k)$-axis  from (2.0,0.0) to (8.0,0.0) (in
units of fm$^{-1}$). Each segment of these contours is discretized with two points
corresponding to the abscissa of the Gauss-Legendre quadrature. Hence,
we take 6 non-resonant continuum shells from $\{p_{3/2}\}_{\rm c}$ both for protons and neutrons.
The real-energy  $p_{1/2}$ continuum shells  are distributed  along the segment [(0,0), (8,0)] which 
is discretized with 6 points for protons and neutrons. 
The real-energy $s_{1/2}$ and $d_{5/2}$ continua are included in the valence space as well. They are distributed
along the real $k$-axis along the segment [(0,0), (3,0)]. We take six  $\{s_{1/2}\}_{\rm c}$ and $\{d_{5/2}\}_{\rm
c}$ discretization points, both for protons and neutrons. The $0s_{1/2}$ poles are not included in the valence
space as they are assumed to be occupied in the core of $^{4}$He. The total number of shells $N_{\rm sh}$ in the
GSM configuration space  is then equal to 50.

\subsection {The three-nucleon case: $J^{\pi}=3/2^{-}$ ground state of $^{7}$Li}
\label{gs7li}

The s.p. basis of $^{7}\rm{Li}$ is generated by the HF potential (calculated separately for protons and neutrons
and for each partial wave). It contains bound s.p. $p_{3/2}$ states at energies $-$5.605\,MeV (neutrons) and
$-$7.098 MeV (protons). These  s.p. states generate  the pole space in the many-body GSM
framework and the reference subspace $A$ in DMRG. As discussed above,
the total number of resonant ($p_{3/2}$) and non-resonant
($\{p_{3/2}\}_{\rm c}$, $\{p_{1/2}\}_{\rm c}$, $\{s_{1/2}\}_{\rm c}$,
$\{d_{5/2}\}_{\rm c}$) shells for protons and neutrons is 50. The
dimension of the Lanczos space spanned by  one valence proton and two
valence neutrons in these 50 shells, i.e., the dimension of the GSM
matrix, is $D$=7,796. The ground state energy $E_{\rm
ex}=(-26.6620,0.2486)$\,MeV  has a non-vanishing, unphysical imaginary part.
This  spurious width comes from the
fact that the discretization along the contours is not precise enough to
effectively fulfill the completeness relation (\ref{Ber_com}). This
problem will be addressed in Sec.~\ref{complet} by increasing the number of points along
the contour. In what follows, we
shall study the convergence of the GSM+DMRG method by varying either the number of eigenvectors $N_{\rm opt}$ kept
during the sweeping phase (see Sec. \ref{gs7lifns}), or the precision $\epsilon$ of the density matrix (see Sec.
\ref{eigsum}).
\subsubsection{DMRG truncation with fixed  $N_{\rm opt}$}
\label{gs7lifns} The number of eigenvectors of $\hat\rho_B$ with the largest nonzero moduli of 
eigenvalues  kept at each iteration
during the warm-up phase is limited to ${N}_{\rm opt}^{(0)}$=26. This number corresponds to the total number of
states $\{n;j_B\}$  in the subspace $B$ that can be coupled with states in $A$ to yield configurations with
$J^{\pi}=3/2^{-}$.

The actual number of eigenvectors kept in the
warm-up phase may be less than
${N}_{\rm opt}^{(0)}$ since most of eigenvectors have vanishing eigenvalues. The
non-resonant continuum shells involved in $B$  are ordered according to the  sequence:
\begin{eqnarray}
\{ \dots \pi p_{3/2}^{(i)},~\nu p_{3/2}^{(i)},~\pi p_{1/2}^{(i)},~\nu p_{1/2}^{(i)},\nonumber \\
\pi s_{1/2}^{(i)},~\nu s_{1/2}^{(i)},~ \pi d_{5/2}^{(i)},~\nu d_{5/2}^{(i)} \dots \},
\label{order}
\end{eqnarray}
where index $i$ denotes the position of scattering shells on their respective contours, beginning with
those  closest to the $k$=0 origin.

\vspace{1cm}
\begin{figure}[hbt]
\begin{center}
\includegraphics[width=0.4\textwidth]{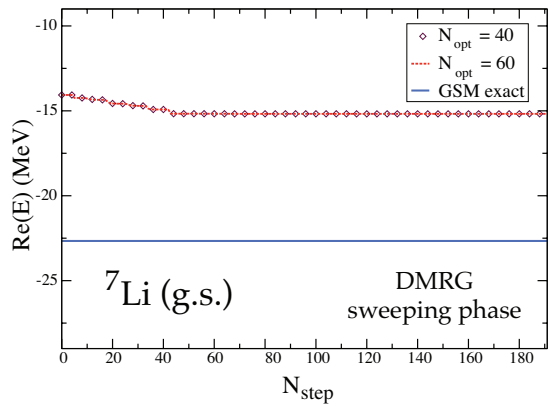}
\caption{The ground-state energy of  $^7$Li as a function of the step number
in the sweeping phase, $N_{\rm step}$. For both values of $N_{\rm opt}$,  the maximum number of
eigenvectors kept during the warm-up phase is $N_{\rm opt}^{(0)}$=26.
The step zero corresponds to  the result obtained at the  end of the  warm-up phase.
 The solid line
marks the exact GSM value  obtained by means of the direct Lanczos diagonalization.}
\label{fig_Li7_re_wrong}
\end{center}
\end{figure}
Figure \ref{fig_Li7_re_wrong} illustrates  the convergence of the GSM+DMRG procedure with
respect to the step number in the sweeping phase for $N_{\rm opt}$=40 and  60.
 The results are identical for both values of
 $N_{\rm opt}$, and the DMRG ground-state
energy converges to the value of
$Re(E_{\rm DMRG})=-$15.176\,MeV. This exceeds by $\sim$7\,MeV
the exact GSM value of  $Re(E_{\rm ex})$=$-$22.662\,MeV, obtained by
the direct Lanczos diagonalization of the GSM Hamiltonian.

Clearly, when applied to the case shown in Fig.~\ref{fig_Li7_re_wrong}, the GSM+DMRG procedure breaks down. The
reason for this failure is not related to a too small value
of $N_{\rm opt}$: indeed, in the case $N_{\rm opt}=60$ the largest 
number of eigenvectors of the density matrix with non-zero eigenvalues is equal to 50.  
The GSM+DMRG iterative method is
trapped in a local minimum, a not uncommon feature of the standard DMRG procedure. A further increase of $N_{\rm
opt}$ does not change the final results which is fully converged. To understand the origin of the failure, let us
analyze the GSM+DMRG wave function in some detail. To this end, the $J^{\pi}=3/2^-$ g.s. wave function of $^7$Li
is decomposed as follows:
\begin{eqnarray}
|\Psi\rangle=c_{p^{3}}|p^{3}\rangle+c_{s^{2}p}|s^{2}p\rangle+c_{d^{2}p}|
d^{2}p\rangle+c_{spd}|spd\rangle,
\end{eqnarray}
 where $c_\nu$'s are the
amplitudes associated with different  three-nucleon GSM configurations
$|\nu\rangle$. The (real parts) of  squared amplitudes $c^2_\nu$   are
shown in Table \ref{table1} for the GSM+GDMRG wave function
corresponding to $N_{\rm opt}$=60 and for the  exact GSM wave function.
\begin{table}
\begin{ruledtabular}
\begin{tabular}{|c|c|c|}
 Conf. & GSM+DMRG &  Exact GSM  \\ \hline
$p^{3}$  & 0.9922 & 0.9239  \\
$s^{2}p$ & 0.0003 & 0.0051 \\
$d^{2}p$ & 0.0075 & 0.0644 \\
$spd$   & 0.0000 & 0.0066
\end{tabular}
\end{ruledtabular}
\caption{\label{table1}
Real part  of the squared shell-model amplitudes in the
$J^{\pi}=3/2^-$ ground state  wave function of $^7$Li obtained in GSM+DMRG ($N_{\rm
opt}=60$) and through the exact Lanczos diagonalization.}
\end{table}
As compared to the exact result, the  $|p^{3}\rangle$ parentage amplitude is
overestimated and the $|spd\rangle$ component  is totally absent in the GSM+DMRG
wave function.
The latter can be understood by observing that (i) only shells with $l$=1
span the reference subspace $A$, and (ii) during the  GSM+DMRG procedure, scattering
shells  are added one by one.
Consequently, when the first positive-parity shell (in our case, a $\pi s_{1/2}$ shell, see
(\ref{order})) is added, the $|spd\rangle$ component
cannot be generated as the first $d_{5/2}$ shell is added only later.
When the
first $d_{5/2}$ non-resonant shell is included, the $|spd\rangle$ configuration
cannot be generated either, because
states with one particle in previously considered $s$-shells are not kept in the
process of optimization due to the parity conservation.
Therefore,   the  $|spd\rangle$ configuration never enters the DMRG wave function; hence,
GSM+DMRG   converges to a wrong solution.
\begin{figure}[hbt]
\begin{center}
\includegraphics[width=0.3\textwidth]{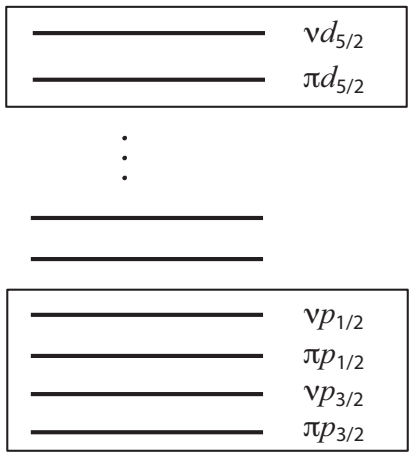}
\caption{Schematic illustration of the reference subspace $A'$ during the
warm-up phase of GSM+DMRG for
$J^{\pi}=3/2^-$ wave functions of $^7$Li.
In addition to the   $\pi(0p_{3/2})$,
$\nu(0p_{3/2})$, $\pi(0p_{1/2})$, and $\nu(0p_{1/2})$ poles, two scattering shells
$\pi d_{5/2}$ and $\nu d_{5/2}$  are now included to generate $|spd\rangle$
configurations during the DMRG procedure. See text for more details.}
\label{medium_d}
\end{center}
\end{figure}
To prevent this pathological behavior, we add to the reference subspace $A$ two 
positive-parity scattering  shells $\pi d_{5/2}$ and $\nu d_{5/2}$ to form a new reference subspace $A'$
(see Fig.~\ref{medium_d}). We arbitrarily choose the last $\pi d_{5/2}$ and $\nu d_{5/2}$ 
shells in the sequence (\ref{order}). (As we shall see later, any 
other positive parity shells can be chosen as well.) The role played  by the additional
positive parity shells is to generate
missing SM couplings in the wave function. The new reference subspace $A'$ is used 
for the construction of the set $\{k_{A'}\otimes i_{B}\}$ and the density matrix $\hat{\rho}^{B}$. 
At each iteration during the warm-up phase, the density matrix 
contains the correlations due to  the  additional positive-parity orbits.
In this way we assure that no possible
couplings are missing during the warm-up phase. We use the same reference
state as before, i.e.,  $|\Psi_{J}^{0}\rangle$ (which is only generated by the resonant shells) 
to select the target state $|\Psi_{J}\rangle$ among the eigenstates of $\hat{H}$.
By the time the first sweep starts, the two shells $(\pi d_{5/2},\nu d_{5/2})$ are 
included in $B$ and
the procedure is carried out  as was described in Sec.~\ref{sec_dm} with the reference subspace $A$.

\vspace{1cm}
\begin{figure}[hbt]
\begin{center}
\includegraphics[width=0.4\textwidth]{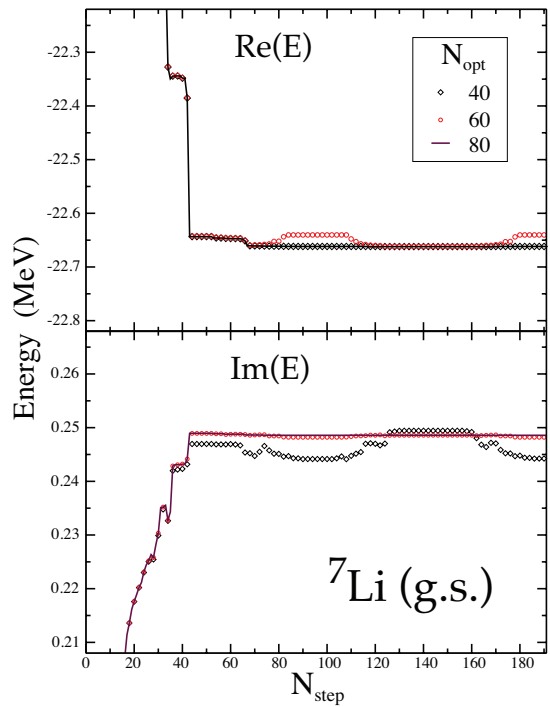}
\caption{The GSM+DMRG ground state energy (real part, top; imaginary part, bottom) of $^7$Li
for $N_{\rm opt}$=40, 60, and 80
as a function of the step number
during the sweeping phase. Two non-resonant continuum shells
($\pi d_{5/2},\nu d_{5/2}$) are included in the reference subspace during
the warm-up phase ($N_{\rm opt}^{(0)}$=26). }
\label{fig_Li7_good_re}
\end{center}
\end{figure}
Real and imaginary parts of the ground state energy, obtained using the
extended reference subspace $A'$ of Fig.~\ref{medium_d}, are plotted in
Fig.~\ref{fig_Li7_good_re} for different values of $N_{\rm opt}$.  For
$N_{\rm opt}$=40, one can see pronounced quasi-periodic oscillations in
both real  and imaginary parts of the energy. These oscillations have
the periodicity of 96 steps corresponding to two consecutive sweeps:
sweep-down and sweep-up, each consisting of 48 steps. The energy
oscillations  rapidly  diminish with increasing $N_{\rm opt}$, and the
calculated energy  $E_{\rm DMRG}$ converges to the GSM benchmark result:
$E_{\rm ex}=(-26.6620,0.2486)$\,MeV. For $N_{\rm opt}$=80, the deviation
from the benchmark result is less than 1\,keV for the real part of the
energy, and less than 0.1\,keV for the corresponding imaginary part.

The rank $d_{\rm H}^{\rm max}$ of largest matrix to be diagonalized grows almost
linearly with $N_{\rm opt}$, from $d_{\rm H}^{\rm max}$=716 for $N_{\rm opt}=40$ ($\sim9.1$ \%
of the dimension $D$ of the GSM matrix)
to $d_{\rm H}^{\rm max}$=1469 ($\sim19$ \% of $D$) for $N_{\rm opt}$=80.
One should keep in mind that ${\rm d}_{\rm H}^{\rm max}$ is almost independent of the
continuum discretization density \cite{Rot2}, i.e., the number of
scattering  shells considered. Hence, the ratio ${\rm d}_{\rm H}^{\rm
max}/D$ decreases rapidly with the  number of valence shells \cite{Rot2}.

\begin{figure}[hbt]
\begin{center}
\includegraphics[width=0.4\textwidth]{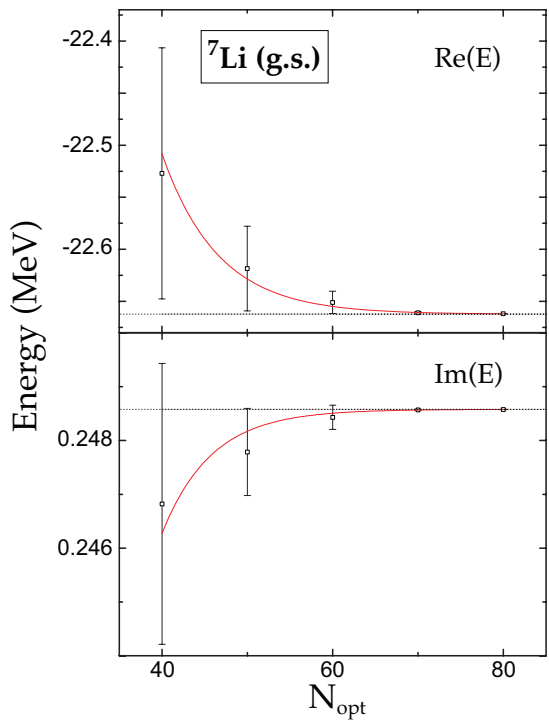}
\caption{Average value of the ground state energy (top: real part; bottom: imaginary part)
of  $^{7}$Li in GSM+DMRG
for different values of $N_{\rm opt}$.
Error bars correspond to the difference between extremum values of
$Re(E_{\rm DMRG})$  (top) and $Im(E_{\rm DMRG})$ (bottom)
at the fourth sweep.
Two scattering  shells  ($\pi d_{5/2},\nu d_{5/2}$) are included in
the reference subspace during the warm-up phase ($N_{\rm opt}^{(0)}$=26).
}
\label{fig_ampl_Li7_N_opt}
\end{center}
\end{figure}
The GSM+DMRG energy averaged over  steps of the fourth sweep,
as well as the minimum and maximum energy value reached during this
sweep, are plotted in Fig. \ref{fig_ampl_Li7_N_opt}
for various $N_{\rm opt}$.
As $N_{\rm opt}$ increases, the amplitude of energy, defined
as the difference between
the maximum and the minimum of real (top) or  imaginary (bottom) part of GSM+DMRG energy
during the fourth sweep,
decreases monotonously. Moreover, with increasing $N_{\rm opt}$, both real
and imaginary parts of the average energy converge  exponentially  to the
exact value. Results of $\chi^2$-analysis
are shown by solid lines in Fig.~\ref{fig_ampl_Li7_N_opt}.
Asymptotic values extracted in this way,
$Re(E^{(\infty)}_{\rm DMRG})=-26.6622\pm 0.0002$\,MeV and
$Im(E_{\rm DMRG}^{(\infty)})=0.248580\pm 0.000004$~MeV,  reproduce
the exact GSM result very well.
The feature of an exponential convergence of  the step-averaged GSM+DMRG
energies may be useful when  estimating   eigenvalues
based  on results obtained with  relatively small  $N_{\rm opt}$.

To illustrate how the generalized variational principle (\ref{genvar1}) works, let 
us consider the energy  with the greatest modulus, $E_{\rm max}$,  calculated
in DMRG during the last sweep. The values of $E_{max}$ and 
$E_{\rm ave}$, the energy averaged during the last sweep  (corresponding
to Fig \ref{fig_ampl_Li7_N_opt}), are
shown in Table~\ref{table_var_N_opt} for different $N_{\rm opt}$. One can
clearly see that the closer the wave function calculated with DMRG is to
the exact wave function as  $N_{\rm opt}$ increases, the larger
$|E_{\rm max}|$ is. Hence, in this case, the modulus of energy reaches a
maximum at the local extremum of the functional $E[\Phi]$ corresponding
to the ground state energy of $^7$Li.
 
The convergence to the exact value is faster by considering $E_{\rm max}$
for each truncation $N_{\rm opt}$ instead of  selecting the average value
$E_{\rm ave}$ (cf. Table \ref{table_var_N_opt}). The real part $Re(E_{\rm max})$ 
converges exponentially  to 
$Re(E^{(\infty)}_{\rm DMRG})=-26.6621\pm 0.0002$\,MeV  while the imaginary
part of $E_{\rm max}$ doesn't follow the exponential behavior.
\begin{table}
\begin{ruledtabular}
\begin{tabular}{|c|c|c|c|c|c|}
 $N_{\rm opt}$. & $|E_{\rm max}|$ & $Re(E_{\rm max})$ &  $Im(E_{\rm max})$ & $Re(E_{\rm ave})$ &  $Im(E_{\rm ave})$  \\ \hline
40 & 22.6489   & -22.6475    & 0.2470    &  -22.5270 & 0.2468  \\
50 & 22.6605   & -22.6591    & 0.2484    & -22.61844 & 0.2478  \\
60 & 22.6631   & -22.6617    & 0.2485    & -22.6510  & 0.2484  \\
70 & 22.6634   & -22.6620    & 0.2486    & 22.6609   & 0.2486  \\
80 & 22.6634   & -22.6620    & 0.2486    & -22.6619  & 0.2486  \\
\end{tabular}
\end{ruledtabular}
\caption{\label{table_var_N_opt}   
Modulus, real and imaginary part of $E_{\rm max}$ defined as the DMRG energy with the greatest
 modulus during the last sweep. The real and imaginary parts of the average energy $E_{\rm ave}$ 
 at the fourth sweep (corresponding to the case presented 
in Fig \ref{fig_ampl_Li7_N_opt}  are also shown for comparison.}
\end{table}
\begin{figure}[hbt]
\begin{center}
\includegraphics[width=0.4\textwidth]{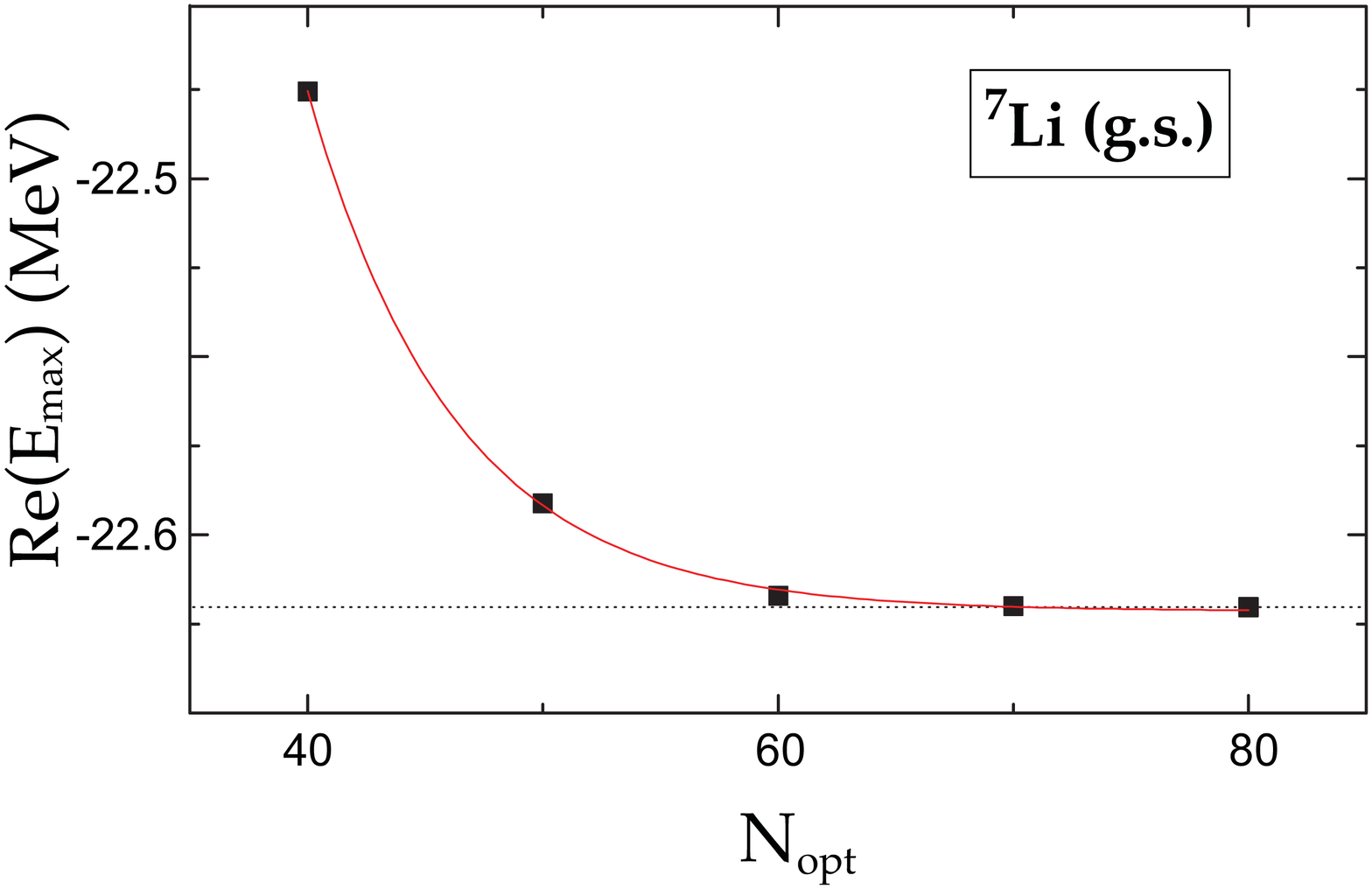}
\caption{Real part of the ground state energy of $^7$Li as a function of
$N_{\rm opt}$. For each $N_{\rm opt}$, the energy  with the largest
modulus during the last sweep  $E_{\rm max}$ is selected. Two scattering
 shells  ($\pi d_{5/2},\nu d_{5/2}$) are included in the reference
subspace during the warm-up phase ($N_{\rm opt}^{(0)}$=26).}
\label{varia_real_N_opt}
\end{center}
\end{figure}

\begin{figure}[hbt]
\begin{center}
\includegraphics[width=0.4\textwidth]{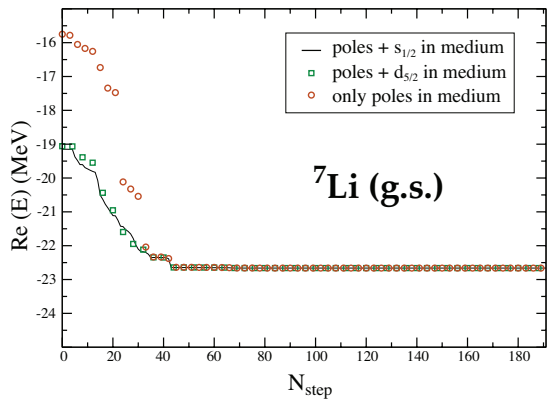}
\caption{
The real part of the ground state energy of   $^7$Li  as a function of $N_{\rm step}$.
The solid line shows results obtained using an
extended reference space spanned on the $(p_{3/2}, p_{1/2})$ poles space and
two ($\pi s_{1/2},\nu s_{1/2}$)
scattering shells. Results depicted by squares have been obtained
using a standard setup for an extended reference subspace
as in Fig.~\protect\ref{fig_Li7_good_re}.
 Open circles show results obtained with a
 reference subspace spanned on the $(p_{3/2}, p_{1/2})$ poles  and
 demanding that at least one state from each $\{n;j_B\}$ family is kept
 during the warm-up phase. The values of  $N_{\rm opt}^{(0)}$=26 and $N_{\rm opt}$=80
were used in all cases.
 See text for more details.}
\label{fig_Li7_P_46s_48add_re}
\end{center}
\end{figure}
The choice of  positive-parity scattering shells to be included in the reference
subspace is somehow arbitrary. The only
important point is that by including both positive and negative parity shells during
the warm-up phase,  one can generate many-body  configurations that would not appear otherwise.
For that reason, one can replace $d_{5/2}$ with  $s_{1/2}$ scattering
shells without changing the outcome of the GSM+DMRG procedure.
To illustrate this, Fig. \ref{fig_Li7_P_46s_48add_re} shows the GSM+DMRG results
with the extended reference subspace $A'$ containing
either two $(\pi s_{1/2},\nu s_{1/2})$ or $(\pi d_{5/2},\nu d_{5/2})$
scattering shells. It is seen that the converged value of the GSM+DMRG energy
is the same in both cases.

A different way to generate the missing components of the
wave function is to demand that at each step during the warm-up phase at
least one state from each $\{n;j_B\}$ family is kept after truncation. 
We take up to $N_{\rm opt}^{(1)}$ eigenvectors of the density matrix
with largest nonzero eigenvalues, where $N_{\rm opt}^{(1)}$ is equal to
the number of different families $\{n;j_B\}$ which contribute to the
GSM+DMRG wave function. If certain families are not represented in this
set of eigenvectors, we add one state for each such family even if the
corresponding eigenvalue equals zero. Hence, the actual number of
vectors kept during the warm-up phase almost always exceeds $N_{\rm
opt}^{(1)}$. Using this additional condition, one may employ a standard
setup for the reference subspace (i.e., $A$  is spanned by s.p.  poles).
Results using this GSM+DMRG strategy are also shown  in Fig.
\ref{fig_Li7_P_46s_48add_re}. The minimal number of states which are
kept in the warm-up phase is, in this case, $N_{\rm opt}^{(1)}$=26. In
spite of a rather different energy at the beginning  of the sweeping
phase, the exact GSM+DMRG energy is reproduced. Moreover, the use of an
extended reference subspace improves convergence. The rank of the
largest matrix to be diagonalized  in this case, $d_{\rm H}^{\rm
max}$=1469, is  independent of the algorithm chosen.

\subsubsection{Truncation governed by the trace of the reduced density matrix}\label{eigsum}

\begin{figure}[hbt]
\begin{center}
\includegraphics[width=0.4\textwidth]{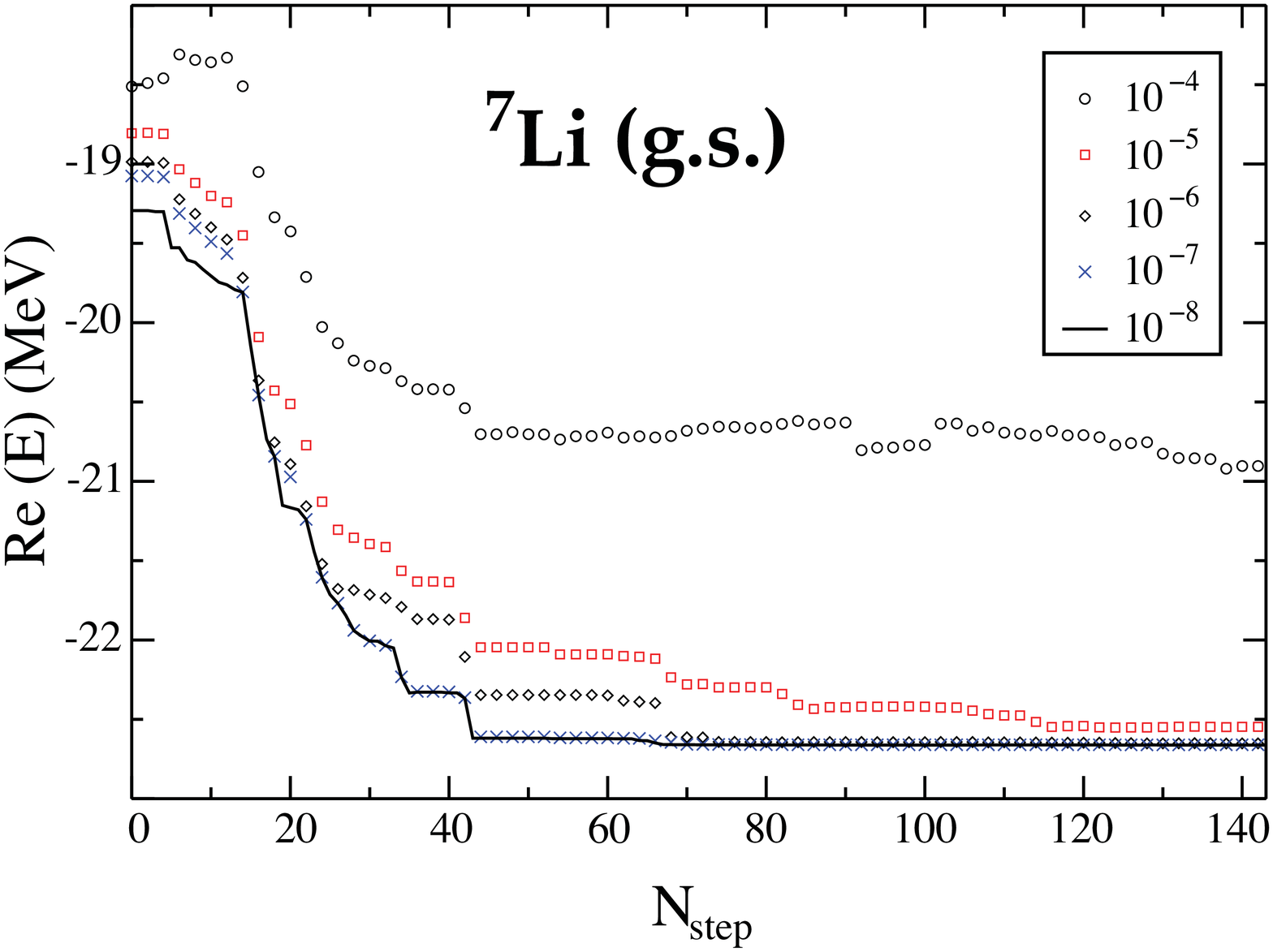}
\caption{The ground state energy (real part) of $^7$Li as a function of
the step number during the sweeping phase for five 
values of the truncation error $\epsilon$ of the reduced density
matrix, see Eq.~(\ref{criterion_den}). Two scattering shells  ($\pi
d_{5/2},\nu d_{5/2}$) are included in the reference subspace during the
warm-up phase.}
\label{fig_en_P_46_density_1e-4_1e-8_add}
\end{center}
\end{figure}

In the  examples described in  Sec.~\ref{gs7lifns},
the maximum number of states $N_{\rm opt}$  is kept fixed at each step of the sweeping
phase. This does not mean that the number of eigenvectors
retained in the sweeping phase is always  constant or equal $N_{\rm opt}$; only the
 eigenvectors of the density
matrix with non-vanishing eigenvalues are kept. In this
section, we shall investigate the GSM+DMRG algorithm in which the number
of states $N_{\rho}$  kept at any step in the sweeping phase depends on
the condition (\ref{criterion_den}) for the trace of the density matrix.
The real part of the $J^\pi=3/2^-_1$ eigenvalue in $^7$Li is shown
in Fig.~\ref{fig_en_P_46_density_1e-4_1e-8_add} for several values of $\epsilon$. As in
the previous examples,
 the reference subspace $A'$ is spanned by the HF
poles and two scattering  shells  $(\pi d_{5/2},\nu d_{5/2})$. In the warm-up phase, we keep $N_{\rm opt}^{(1)}$
eigenvectors of the density matrix (up to 26) and additionally require that at least one state of each $\{n;j_B\}$
family is retained.  As before, $N_{\rm opt}^{(1)}$ is equal to the number of different $\{n;j_B\}$ families so
the total number of eigenvectors kept at each iteration step is greater or equal to
 $N_{\rm opt}^{(1)}$. For low-precision
calculations ($\epsilon=10^{-4}$), the resulting energy
oscillates  and approaches a value which deviates from the correct result
 by $\sim$1.9\,MeV. The
amplitude of  oscillations as a function of $N_{\rm step}$
quickly decreases  with decreasing $\epsilon$. For
$\epsilon=10^{-8}$, the precision of the converged GSM+DMRG energy value is $\simeq$
0.2\,keV  for both real and imaginary parts.
\vspace{1cm}

Obviously, the dimension of the largest matrix to be  diagonalized
depends on the required truncation error  $\epsilon$.
In the studied case, $d_{\rm H}^{\rm max}$ changes from 273 for
$\epsilon=10^{-4}$ to 1327 for $\epsilon=10^{-8}$ with the average number of
vectors kept during the sweeping phase increasing  from $\sim$15 to $\sim$46.
In the truncation scenario
with fixed $N_{\rm opt}$, the number of saved vectors, averaged over one
sweep,   is $\sim$59 for $N_{\rm
opt}$=80. In general, for the same precision of  GSM+DMRG
energies, the average number of vectors kept during the sweeping phase
is smaller if the truncation is done dynamically according to the trace
of density matrix than by fixing the maximum number of eigenvectors
$N_{\rm opt}$.
\begin{figure}[hbt]
\begin{center}
\includegraphics[width=0.4\textwidth]{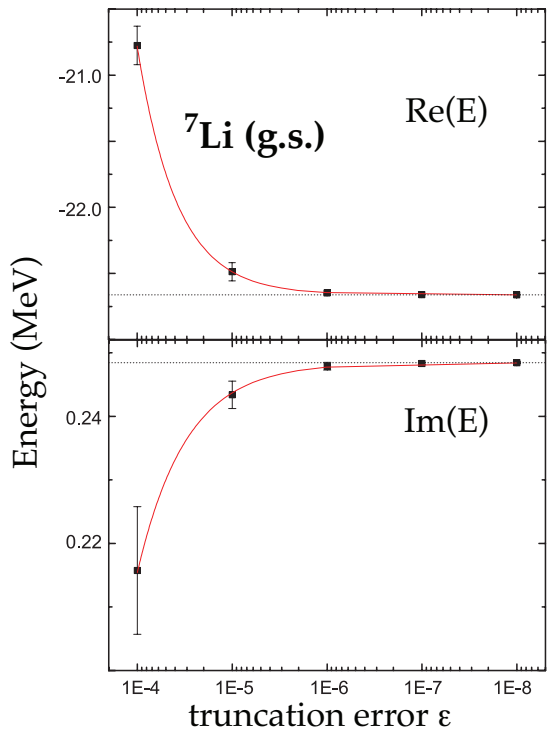}
\caption{The ground state average energy (top: real part; bottom: imaginary part) of $^{7}Li$  as a
function of the truncation error $\epsilon$ of the reduced density matrix.
Two scattering shells  ($\pi
d_{5/2},\nu d_{5/2}$) are included in the reference subspace during the
warm-up phase. Error bars correspond to a difference between the
extremum values of $Re(E_{\rm DMRG})$
at the third sweep.
The
solid line shows the results of $\chi^2$ analysis, assuming power-law convergence (\ref{chi2}).}
\label{fig_ampl_fixed_den}
\end{center}
\end{figure}
The GSM+DMRG energy averaged over steps of the third sweep as well as
the minimum and maximum energy  reached during this sweep are
plotted in Fig.~\ref{fig_ampl_fixed_den}
as a function  of $\epsilon$. The
GSM+DMRG error, i.e., the energy difference with respect to the exact GSM
result, decreases fast with decreasing  $\epsilon$.
 The real and imaginary parts of DMRG energy satisfy to a good
approximation the power law 
\begin{eqnarray}\label{chi2}
Re\left(E_{\rm DMRG}\right)=Re\left(E_{\rm ex}\right)+\alpha \epsilon^{\beta}
\end{eqnarray}
proposed  in Ref.~\cite{Leg03} to control the accuracy of the DMRG method for Hermitian problems.
The results of a $\chi^2$-fit to Eq.~(\ref{chi2})  are shown in Fig.~\ref{fig_ampl_fixed_den}.
The asymptotic values extracted in this way are 
$Re(E^{(\infty)}_{\rm DMRG})=-26.66192\pm 2\cdot 10^{-5}$~MeV and
$Im(E_{\rm DMRG}^{(\infty)})=0.24844\pm 3\cdot 10^{-5}$~MeV 
and agree very well with the exact GSM energy.

\begin{figure}[hbt]
\begin{center}
\includegraphics[width=0.4\textwidth]{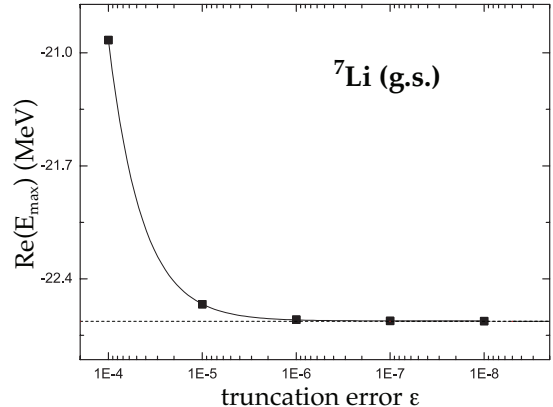}
\caption{Real part of the ground state energy of $^7$Li  as a function of $\epsilon$.
At each $\epsilon$,  the energy is selected according to
the greatest modulus during the last sweep.}
\label{varia_real_den}
\end{center}
\end{figure}

In  Table \ref{table_var_density} we compare the average complex energy $E_{\rm ave}$
at the third sweep for different values of $\epsilon$ (corresponding to Fig \ref{fig_ampl_fixed_den}
and the complex energy $E_{\rm max}$ (the energy
with the greatest modulus during the last sweep). As in the previous case where $N_{opt}$ was fixed, the modulus
of $E_{\rm max}$  reaches a maximum when $E_{\rm max}$ is equal to the exact GSM energy. 
As  can be seen in Fig.~\ref{varia_real_den},
the real part of $E_{\rm max}$ exhibits a power-law behavior
 with an extrapolated
value equal to $Re(E^{(\infty)}_{\rm DMRG})=-26.660.5\pm 0.0012$\,MeV.
\begin{table}
\begin{ruledtabular}
\begin{tabular}{|c|c|c|c|c|c|}
 $\epsilon$. &  $|E_{\rm max}|$ &  $Re(E_{\rm max})$ & $Im(E_{\rm max})$ & $Re(E_{\rm ave})$ & $Im(E_{\rm ave})$ \\ \hline
$10^{-4}$ &  20.9221    & -20.9209    &  0.2240    & -20.7751    & 0.2157    \\
$10^{-5}$ &  22.5575    & -22.5562    &  0.2416    & -22.4870    & 0.2434    \\
$10^{-6}$ &  22.6532    &  -22.6519   &  0.2485    & -22.6474    & 0.2479    \\
$10^{-7}$ &  22.6621    &  -22.6607   &  0.2481    & -22.6602    & 0.2483    \\
$10^{-8}$ & 22.6632     &  -22.6618   &  0.2486    & -22.6618    & 0.2484 \\
\end{tabular}
\end{ruledtabular}
\caption{\label{table_var_density}   
Modulus, real and imaginary part of  $E_{\rm max}$ defined as  the
energy having the greatest modulus during the last sweep. The real and imaginary
parts of the average energy $E_{average}$ at the fourth sweep
(see Fig \ref{fig_ampl_fixed_den}) are also shown for comparison.}
\end{table}

\subsubsection{Treatment of spurious width} \label{complet}
As  mentioned in Sec.~\ref{gs7li}, the imaginary part of the GSM 
 energy $E_{\rm ex}=(-26.6620, 0.2486)$\,MeV is non-physical for it has
a negative  width. This is due to
the fact that the contour discretization is  not sufficiently precise  to 
  guarantee the completeness relation (\ref{Ber_com}). This
spuriosity can be taken care of by increasing the number of points
along the integration contour. In the largest calculation we have done for the
ground state of $^{7}$Li, we took 67 points along the
contour $L_{\pi;p_{3/2}}^{+}$, 24 along $L_{\pi;p_{3/2}}^{+}$, and 12
points along the contours $L_{\pi;s_{1/2}}^{+}$ and
$L_{\pi;d_{5/2}}^{+}$. The neutron valence space is the same as the proton space
except for the contour
$L_{\nu;p_{3/2}}^{+}$ where 74 points are considered. The model
space corresponds to  239 shells and the  dimension of the
ground state $J^{\pi}=3/2^{-}$ GSM Hamiltonian matrix is 1,459,728.

In order to perform calculations within this huge valence space, we have
developed a parallel version of the DMRG code. At each step during the
DMRG procedure, calculations of the matrix elements of the suboperators
(\ref{suboper})  and  hamiltonian (\ref{hamiltonian})  are
distributed among the processors. Our calculations were carried out on the CRAY XT4
 Jaguar supercomputer at the Oak Ridge National Laboratory.

\begin{figure}[hbt]
\begin{center}
\includegraphics[width=0.4\textwidth]{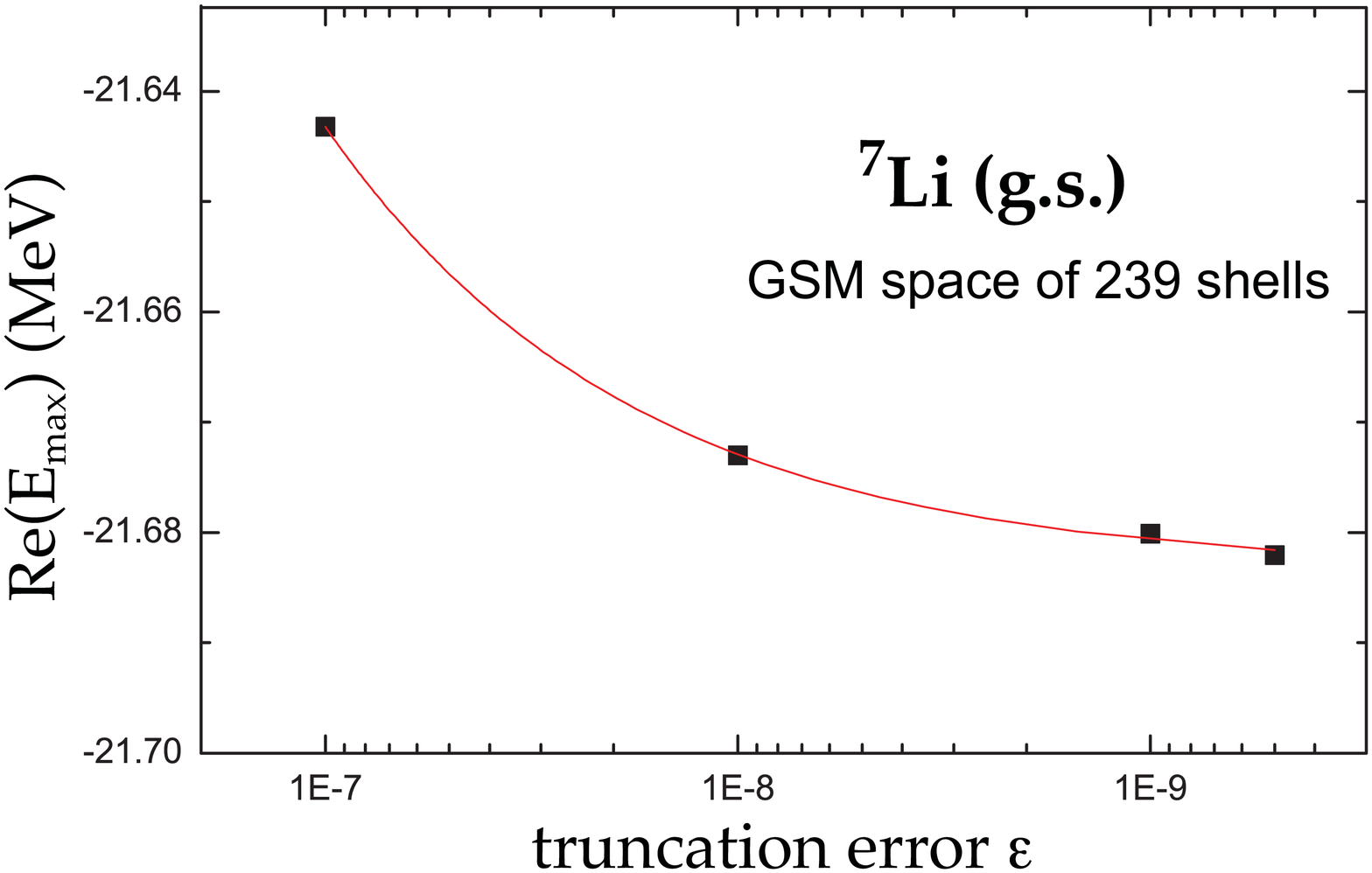}
\caption{Real part of the ground state energy of $^7$Li in a model space made of 239 shells.
At each $\epsilon$,  the energy is selected according to
the greatest modulus during the third sweep. See text for details.}
\label{Lithium239}
\end{center}
\end{figure}

The real part of the ground state energy  and the fit
according to the relation (\ref{chi2}) are plotted in Fig \ref{Lithium239}. For each $\epsilon$ the 
energy $E_{\rm max}$ with the greatest modulus during the third sweep  is considered. The extrapolated 
value is $Re(E^{(\infty)}_{\rm DMRG})=-21.6834\pm 0.0010$\,MeV. The real part of $E_{max}$
for $\epsilon=5\cdot 10^{-10}$ is -21.6820 \,MeV and the amplitude during the last sweep is
2.275 \,keV; hence, convergence has almost been  reached. Here, the largest matrix
 has a dimension 3,348. The imaginary
part (which does not follow the power law behavior)  varies from 0.00100
 \,MeV at $\epsilon=10^{-7}$ to 0.00075 \,MeV
at $\epsilon=5\cdot 10^{-10}$ (its amplitude during the last sweep is 0.065 \,keV).
This example nicely demonstrates the validity of the many-body  completeness relation in GSM.

\subsection {The four-nucleon case: $J^{\pi}=2^{+}$  ground state of $^{8}$Li}
\label{gs8li}

The GSM Hamiltonian for $^8$Li is the same as for $^{7}$Li, the only modification being the 
change of $T$=0 couplings  in Eq.~(\ref{param_V}) due to the different
number of neutrons $N_v$=3. The HF procedure yields two bound s.p.
states: $e_{p_{3/2}}=-8.556$\,MeV  and $e_{p_{3/2}}=-12.788$\,MeV, for
neutrons and protons, respectively. Shells of the non-resonant continuum
are distributed in the complex $k$-plane using the same contours and the
same discretization scheme as in the $^{7}$Li case. The dimension of the
Lanczos space spanned by one valence proton and three valence neutrons
in 50 shells, i.e., the dimension of the GSM matrix in $^8$Li, is
$D$=170,198. The GSM+DMRG results presented in this section are
obtained using the truncation criterion (\ref{criterion_den}). As
discussed in Sec.~\ref{eigsum}, this criterion is somewhat more
efficient than the condition based on fixing the maximum number of
eigenvectors $N_{\rm opt}$.
The truncation method employed in the warm-up phase follows that of
Sec.~\ref{gs7lifns}. The reference subspace $A'$ is
spanned by  the pole states and two scattering shells ($\pi d_{5/2},\nu d_{5/2}$).
We take up to 50 eigenvectors of the density matrix with the largest
nonzero eigenvalues.  If certain $\{n;j_B\}$ families are not represented in
this set of eigenvectors, we add one state for each such family even if
the corresponding eigenvalue equals zero. In the sweeping phase, we follow
the truncation strategy of  Sec.~\ref{eigsum}.

\vspace{1cm}
\begin{figure}[hbt]
\begin{center}
\includegraphics[width=0.4\textwidth]{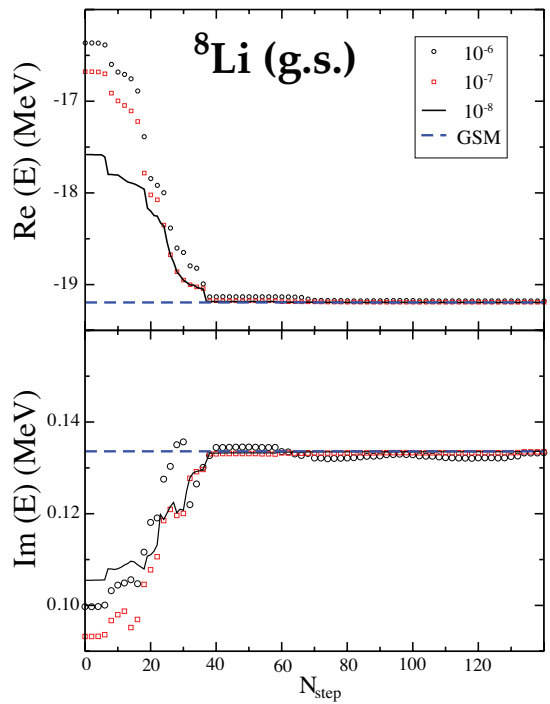}
\caption{The real part of the ground state energy of $^8$Li as a function of step number
during the sweeping phase. Results are shown for different values of the
truncation error $\epsilon$. Two non-resonant continuum shells  ($\pi
d_{5/2},\nu d_{5/2}$) are included in the reference subspace during the
warm-up phase.}
\label{fig_en_density_Li8_re}
\end{center}
\end{figure}
\begin{figure}[hbt]
\begin{center}
\includegraphics[width=0.4\textwidth]{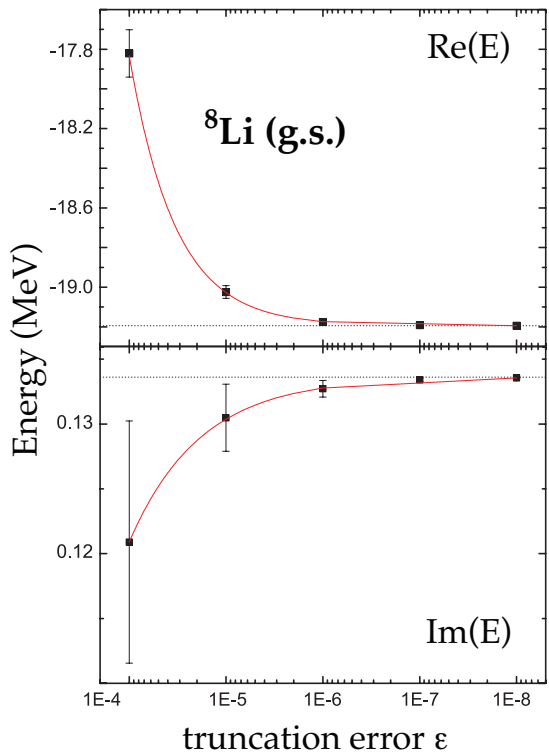}
\caption{Similar as in Fig.~\protect\ref{fig_ampl_fixed_den}
except  for the ground state average energy of $^8$Li.}
\label{fig_ampli_Li8_re}
\end{center}
\end{figure}

Figure~\ref{fig_en_density_Li8_re} shows the DMRG+GSM results for $^8$Li
 for three values of $\epsilon$.  The exact energy of the
 $2^+_1$  state obtained by
the direct Lanczos diagonalization of the GSM matrix is  $E_{\rm
ex}=(-19.19451,0.13361)$\,MeV.
The corresponding energies averaged over
the third sweep are plotted in Fig.~\ref{fig_ampli_Li8_re}.
For $\epsilon=10^{-4}$, the largest matrix to be
diagonalized has a rank $d_{\rm H}^{\rm max}$=1,446 ($\sim0.8$\% of $D$).
 For $\epsilon=10^{-8}$, one obtains
$E_{\rm DMRG}=(-19.19415,0.13355)$\,MeV, i.e. the real part of the
GSM+DMRG energy  deviates only by 0.4\,keV from  the exact value, while for the
 imaginary part, the deviation is less than
0.06\,keV. The largest matrix to be diagonalized in this case has
a rank ${\rm d}_{\rm H}^{\rm max}$=20,535 ($\sim 12$ \% of $D$).
%

\section{Conclusions}\label{sec_sum}

This work describes the first application of the DMRG method to two-fluid, open  many-fermion systems represented
by complex-symmetric Hamiltonians. Calculations were carried out for proton-neutron systems $^7$Li (three-nucleon
problem) and  $^8$Li (four-nucleon problem). As compared to our previous work \cite{Rot2}, two significant
improvements of the GSM+DMRG technique have been made. The first improvement concerns the recognition of the
appropriate target state in the warm-up phase. The second development relates to the truncation strategy in the
sweeping phase.

There are situations in which the DMRG procedure yields a fully
converged but incorrect solution. In order to understand and prevent
this pathological behavior, we studied a necessary and sufficient
condition for the GSM+DMRG method to yield a correct eigenvalue.
The essential condition is to assure that all possible couplings in
the many-body wave function, allowed by the symmetries of the problem
and the configuration space, are present  in the warm-up phase. We
propose different strategies to guarantee this crucial requirement.

Two truncation schemes for the number of retained vectors in the
sweeping phase of DMRG were investigated: the fixed-$N_{\rm opt}$
scheme   (Sec.~\ref{gs7lifns}) and the dynamic truncation
(Sec.~\ref{eigsum}). We conclude that the two  strategies are to a
large extent  equivalent; they  both exhibit the excellent convergence
properties to the benchmark GSM result. In both cases, one finds the
quasi-periodic oscillations of GSM+DMRG energy as a function of $N_{\rm
step}$ with extensive plateaux.

The  GSM+DMRG energy averaged over one sweep exhibits excellent exponential convergence with  $N_{\rm opt}$ which
allows to deduce the asymptotic value with good precision. Also, $E_{\rm DMRG}$ exhibits excellent convergence
as a function of the truncation error $\epsilon$. This feature makes it possible to control the accuracy of
GSM+DMRG calculations. 

The dynamic truncation, fixing a condition on the trace of the reduced density matrix, yields results of similar
accuracy for $^7$Li and $^8$Li, i.e., for systems having very different configuration spaces. This offers a
possibility to compare the convergence in different quantal systems at the same value of $\epsilon$.

The encouraging features of the proposed GSM+DMRG approach open the
window for systematic and high-precision studies of complex, 
weakly bound nuclei, such as halo systems, which require large
configuration spaces involving s.p. states of different parities (both
in the pole space and in the scattering space). Generally, the
improvements of the DMRG approach proposed in this work can be of
interest in the context of other multiparticle open quantum systems, as
well as for other DMRG calculations involving non-Hermitian
Hamiltonians.

\section{Acknowledgements}
We thank Gaute Hagen for usefull discussions. This work was supported by the U.S. Department of Energy under Contract
Nos. DE-FG02-96ER40963 (University of Tennessee), DE-AC05-00OR22725 with
UT-Battelle, LLC (Oak Ridge National Laboratory), and DE-FG05-87ER40361
(Joint Institute for Heavy Ion Research), by the Spanish DGI under grant
No. FIS2006-12783-c03-01, and by the CICYT-IN2P3 cooperation.
Computational resources were provided by the National Center for
Computational Sciences at Oak Ridge and the National Energy Research
Scientific Computing Facility.


\begin{thebibliography}{100}
\bibitem{Jac_d} J. Dobaczewski and W. Nazarewicz, Phil. Trans. R. Soc. Lond. A {\bf 356}, 2007 (1998).
\bibitem{Jac_o} J. Oko{\l}owicz, M. P{\l}oszajczak, and I. Rotter, Phys. Rep. {\bf 374}, 271 (2003).
\bibitem{ppnp} J. Dobaczewski, N. Michel, W. Nazarewicz, M. P{\l}oszajczak, and
J. Rotureau, Prog. Part. Nucl. Phys. {\bf 59}, 432 (2007).
\bibitem{Kar} K. Bennaceur, F. Nowacki, J. Oko{\l}owicz and M. P{\l}oszajczak, Nucl. Phys. A {\bf 651},289 (1999);\\
K. Bennaceur, F. Nowacki, J. Oko{\l}owicz, and M. P{\l}oszajczak, Nucl. Phys. A {\bf 671}, 203 (2000).
\bibitem {Rot} J. Rotureau, J. Oko{\l}owicz, and M. P{\l}oszajczak, Phys. Rev. Lett. {\bf 95}, 042503 (2005);\\
J. Rotureau, J. Oko{\l}owicz, and M. P{\l}oszajczak, Nucl. Phys. A {\bf 767}, 13 (2006).
\bibitem {Volya} A. Volya and V. Zelevinsky, Phys. Rev. C {\bf 74}, 064314 (2006).
\bibitem{Mic02a} N. Michel, W. Nazarewicz, M. P{\l}oszajczak, and K. Bennaceur, Phys. Rev. Lett. {\bf 89}, 042502 (2002);\\
 N. Michel, W. Nazarewicz, M. P{\l}oszajczak, and J. Oko{\l}owicz, Phys. Rev. C {\bf 67}, 054311 (2003).
 \bibitem{Bet02}
R. Id Betan, R.J. Liotta, N. Sandulescu, and T. Vertse, Phys. Rev. Lett. {\bf 89}, 042501 (2002);\\
R. Id Betan, R.J. Liotta, N. Sandulescu, and T. Vertse, Phys. Rev. C {\bf 67}, 014322 (2003).
\bibitem{Bet04} R. Id Betan, R.J. Liotta, N. Sandulescu, and T. Vertse, Phys. Lett. B {\bf 584}, 48 (2004).
\bibitem{Mic06} N. Michel, W. Nazarewicz, M. P{\l}oszajczak, and J. Rotureau, Phys. Rev. C {\bf 74}, 054305 (2006).
\bibitem{Hag06}  G. Hagen, M. Hjorth-Jensen, and N. Michel,  Phys. Rev. C \textbf{73}, 064307 (2006).
\bibitem{Berggren} T. Berggren, Nucl. Phys. A {\bf 109}, 265 (1968);\\
T. Berggren and P. Lind, Phys. Rev. C {\bf 47}, 768 (1993).
\bibitem{dmrg1} S.R. White, Phys. Rev. Lett. {\bf 69}, 2363 (1992);
Phys. Rev. B {\bf 48}, 10345 (1993).
\bibitem{Rev1} J. Dukelsky and S. Pittel, Rep. Prog. Phys. {\bf 67}, 513 (2004).
\bibitem{Rev2} U. Schollw\"ock, Rev. Mod. Phys. {\bf 77},  259 (2005).
\bibitem{Rev3} K. Hallberg, Adv. Phys. {\bf 55}, 477 (2006).
\bibitem{pap2} S.R. White and R. L. Martin, J. Chem Phys. {\bf 110}, 4127 (1999).
\bibitem{chan} D. Ghosh, J. Hachmann, T. Yanai, and G.K. Chan, ÊJ. Chem. Phys. {\bf 128} 144117 (2008).
\bibitem{pap3} J. Dukelsky and G. Sierra, Phys. Rev. Lett. {\bf 83}, 172 (1999).
\bibitem{delft} D. Gobert, U. Schollw\"ock, and J. von Delft, Eur. Phys. J. B {\bf 38}, 501 (2004).
\bibitem{pap4} N. Shibata and D. Yoshioka, Phys. Rev. Lett. {\bf 86}, 5755 (2001).
\bibitem{weiss} Y. Weiss and R. Berkovits, Solid ÊState Commun. {\bf 145}, 585 (2008).
\bibitem{pap1} J. Dukelsky, S. Pittel, S.S. Dimitrova, and M.V. Stoitsov, Phys. Rev. C {\bf 65}, 054319 (2002).
\bibitem{Fei} A.E. Feiguin, E. Rezayi, C. Nayak, and S. Das Sarma, ÊPhys. Rev. Lett. {\bf 100}, 166803 (2008).
\bibitem{pap6} E. Carlon, M. Henkel, and U. Schollw\"ock, Eur. J. Phys.
B {\bf 12}, 99 (1999).
\bibitem{pap5} T. Papenbrock and D.J. Dean, J. Phys. G {\bf 31}, S1377 (2005).
\bibitem{pitsan} S. Pittel and N. Sandulescu, Phys. Rev. C {\bf 73}, 014301 (R) (2006).
\bibitem{pitsan1} S. Pittel, B. Thakur,  and N. Sandulescu, arXiv:0808.1303 (2008).
\bibitem{Rot2}  J. Rotureau, , N. Michel, W. Nazarewicz, M. P{\l}oszajczak, and J. Dukelsky,  Phys. Rev. Lett. {\bf 97}, 110603 (2006).
\bibitem{moisey1} N. Moiseyev, P.R. Certain, and F. Weinhold, Mol. Phys. {\bf 36}, 1613 (1978). 
\bibitem{moisey} N. Moiseyev,  Phys. Rep. {\bf 302}, 212 (1998).
\bibitem{moisey2} N. Moiseyev, Chem. Phys. Lett. {\bf 99}, 364 (1983).
\bibitem{Leg03}
\"{O}. Legeza, J. R\"{o}der, and B.A. Hess, Phys. Rev. B {\bf 67}, 125114 (2003).
\bibitem{Suzuki}
Y. Suzuki and Wang Jing Ju, Phys. Rev. C {\bf 41}, 736 (1990).
\bibitem{Mic04} N. Michel, W. Nazarewicz, and M. P{\l}oszajczak, Phys. Rev. C {\bf 70}, 064313 (2004).
\bibitem{Mcdmrg} I. McCulloch and M. Gulacsi, Europhys. Lett. {\bf 57}, 852
(2002).
\end{thebibliography}
\end{document}